\documentclass[aps,pre,showpacs]{revtex4}
\usepackage{graphicx}
\usepackage{dcolumn}% Align table columns on decimal point
\usepackage{bm}% bold math

\begin{document}
%
%\title{New methods for the numerical solution of Maxwell's
%equations\footnote{To be presented at the International Computational Accelerator
%Physics Conference 2002}}
\title{Numerical methods for solving the time-dependent
Maxwell equations\footnote{Invited talk presented at the International Computational Accelerator
Physics Conference 2002}}
\author{%
H. De Raedt\footnote{E-mail: deraedt@phys.rug.nl},
J.S. Kole\footnote{E-mail: j.s.kole@phys.rug.nl},
K.F.L. Michielsen\footnote{E-mail: kristel@phys.rug.nl},
M.T. Figge\footnote{E-mail: m.t.figge@phys.rug.nl}
}
\affiliation{%
Applied Physics - Computational Physics\footnote{http://www.compphys.org},
Materials Science Centre\\
University of Groningen, Nijenborgh 4\\
NL-9747 AG Groningen, The Netherlands
}
\date{\today}
\begin{abstract}%
We review some recent developments in numerical algorithms to solve the
time-dependent Maxwell equations for systems with spatially varying
permittivity and permeability.
We show that the Suzuki product-formula approach can be used to construct
a family of unconditionally stable algorithms, the conventional Yee algorithm, and
two new variants of the Yee algorithm that do not require the use of the staggered-in-time grid.
We also consider a one-step algorithm, based on the Chebyshev polynomial expansion, and
compare the computational efficiency of the one-step,
the Yee-type, the alternating-direction-implicit, and the unconditionally stable algorithms.
For applications where the long-time behavior
is of main interest, we find that the one-step algorithm may be orders of magnitude more efficient
than present multiple time-step, finite-difference time-domain algorithms.
\end{abstract}


\maketitle

%%%%%%%%%%%%%%%%%%%%%%%%%%%%%%%%%%%%%%%%%%%%%%%%%%%%%%%%%%%%%%%%%%%%%%%%%%%%
%
\newcommand{\threevec}[3]{\left(\begin{array}{c}%
 #1 \\ #2 \\ #3 \end{array}\right)}
\newcommand{\ve}{\varepsilon}
\newcommand{\sve}{\sqrt{\varepsilon}}
\newcommand{\smu}{\sqrt{\mu}}
\def\bE{{\mathbf{E}}}
\def\bH{{\mathbf{H}}}
\def\bD{{\mathbf{D}}}
\def\bB{{\mathbf{B}}}
\def\bX{{\mathbf{X}}}
\def\bY{{\mathbf{Y}}}
\def\bZ{{\mathbf{Z}}}
\def\bJ{{\mathbf{J}}}
\def\bj{{\mathbf{s}}}
\def\bk{{\mathbf{k}}}
\def\be{{\mathbf{e}}}
\def\br{{\mathbf{r}}}
\def\bPsi{{\mathbf{\Psi}}}
\def\bPhi{{\mathbf{\Phi}}}
\def\bsigma{{\mathbf{\sigma}}}
\def\bXi{{\mathbf{\Xi}}}
\def\Two{$\widetilde U_2$ }
\def\Tfour{$\widetilde U_4$ }
\def\bzero{{\mathbf{0}}}
\newcommand{\php}{\phantom{+}}
\newcommand{\pht}{\phantom{T}}
\newcommand{\dd}[1]{\frac{\partial}{\partial #1}}
%

%-------------------------------------------------------------
%
\section{Introduction}\label{sec1}
%
%-------------------------------------------------------------
The Maxwell equations describe the evolution of electromagnetic (EM) fields
in space and time~\cite{BornWolf}.
They apply to a wide range of different physical situations and play an
important role in a large number of engineering applications.
In many cases, numerical methods are required to solve Maxwell's equations~\cite{Taflove,Kunz}.
A well-known class of algorithms is based on a method proposed by Yee~\cite{Yee66}.
This finite-difference time-domain (FDTD) approach owes its popularity mainly due to its flexibility and
speed while at the same time it is easy to implement~\cite{Taflove,Kunz}.

A limitation of Yee-based FDTD techniques is that their stability is
conditional, depending on the mesh size of the spatial discretization and
the time step of the time integration~\cite{Taflove,Kunz}.
Furthermore, in practice, the amount of computational work required to solve the time-dependent Maxwell equations
by present FDTD techniques~\cite{Taflove,Kunz,Website,Zheng1,Namiki,Zheng2,Garcia,Harsh00,Kole01,Kole02}
prohibits applications to a class of important fields such as
bioelectromagnetics and VLSI design~\cite{Taflove,Gandi,Houshmand}.
The basic reason for this is that the time step in the FDTD calculation has to be relatively small in
order to maintain stability and a reasonable degree of accuracy in the time integration.
Thus, the search for new algorithms that solve the Maxwell equation focuses on
removing the conditional stability of FDTD methods and on improving
the accuracy/efficiency of the algorithms.

A systematic approach to construct unconditionally stable algorithms
is to employ a Suzuki product-formula~\cite{Suzuki77}
to approximate the time evolution operator~\cite{DeRaedt87}.
In the case of EM fields, the latter is the matrix exponential of a skew-symmetric matrix
and the approximations take the form of products of
orthogonal transformations~\cite{Kole01,Kole02}.
The resulting numerical algorithms are
unconditionally stable by construction~\cite{DeRaedt87,Smith85}.

The spectral-domain split-operator technique proposed in Ref.\cite{Harsh00}
is one of the many forms that results from the use
of the Lie-Trotter-Suzuki product formulas~\cite{Suzuki77}.
This technique makes use of Fast Fourier Transforms
to compute the matrix exponentials of the displacement operators.
The choice made in Ref.~\cite{Harsh00} yields an approximation to the time-evolution operator that
is no longer orthogonal and hence unconditional stability
is not automatically guaranteed~\cite{COMMENT1}.
In contrast, the methodology proposed in Refs.~\cite{Kole01,Kole02} yields efficient, explicit,
unconditionally stable schemes that operate on the EM fields defined on the real space grid only.
These algorithms %do not suffer from wrap-around effects~\cite{Harsh00}, and
naturally allow for the spatial variations in both the permittivity and the permeability.

The Suzuki product-formula approach also provides a unified framework to construct and analyse other
time stepping algorithms~\cite{DeRaedt87,Horvath}. To illustrate this point we show that the conventional Yee algorithm
and the alternating-direction-implicit (ADI) time-stepping algorithms~\cite{Zheng1,Namiki,Zheng2,Garcia,Horvath}
fit into this framework. Furthermore we propose new variants of the Yee algorithm.

Another route to improve upon the accuracy/efficiency of time-integration schemes
is to make use of the Chebyshev polynomial expansions of
the matrix exponential~\cite{TAL-EZER0,TAL-EZER,LEFOR,Iitaka01,SILVER,LOH}.
In this paper we also discuss a one-step algorithm, based on Chebyshev polynomials,
to solve the time-dependent Maxwell equations for (very) large time steps~\cite{hdrcheb0,hdrcheb1}.

The main purpose of this paper is to review the basic ideas behind the recent developments
in numerical algorithms to solve the time-dependent Maxwell equations and to compare
the virtues and shortcomings of the different methods.
The plan of the paper is as follows.
In Section \ref{sec2} we briefly discuss the basic physical symmetries of the time-dependent Maxwell equations.
The general framework to construct time-integration algorithms is laid out in Section \ref{sec3}.
We also pay attention to the numerical treatment of the current source term.
In Section \ref{sec4} we use the simplest case of the time-dependent Maxwell equations
to illustrate how the various algorithms can be implemented.
We explicitly show how the conventional Yee algorithm naturally fits into this framework and, by minor modification,
construct second-order and fourth-order time-accurate schemes that do not require the use
of staggered-in-time fields, nor extra memory to store intermediate results.
Then we recall the steps to construct the unconditionally stable algorithms proposed
in Ref.~\cite{Kole01,Kole02} and analyse a modification to improve the time-integration accuracy.
Finally we discuss the implementation of the ADI and one-step algorithms.
A discussion of the results of numerical experiments and our conclusions are given
in Section \ref{sec5} and \ref{sec6} respectively.

%-------------------------------------------------------------
%
\section{Theory}\label{sec2}
%
%-------------------------------------------------------------
We consider EM fields in linear, isotropic, nondispersive and lossless materials.
The time evolution of EM fields in these systems is governed by the time-dependent
Maxwell equations~\cite{BornWolf}.
Some important physical symmetries of the Maxwell equations can be made explicit by introducing
the fields
\begin{equation}
\bX(t)\equiv\sqrt{\mu}\,\bH(t)\quad {\rm and} \quad
\bY(t)\equiv\sqrt{\varepsilon}\,\bE(t)\,. \label{EY}
\end{equation}
Here, $\bH(t)=(H_x(\br,t),H_y(\br,t),H_z(\br,t))^T$ denotes the magnetic
and $\bE(t)=(E_x(\br,t),E_y(\br,t),E_z(\br,t))^T$
the electric field vector, while $\mu=\mu(\br)$ and $\ve=\ve(\br)$ denote,
respectively, the permeability and the permittivity.
In the absence of electric charges, Maxwell's curl equations~\cite{Taflove}
read
\begin{equation}
\dd{t}
\left(
\begin{array}{c}
\bX(t) \\
\bY(t)
\end{array}
\right)
={\mathcal H}
\left(\begin{array}{c} \bX(t) \\\bY(t) \end{array} \right)
- \frac{1}{\sqrt{\ve}} \left(\begin{array}{c} 0 \\ \bJ(t)   \end{array} \right),
\label{TDMEJ}
\end{equation}
where $\bJ(t)=(J_x(\br,t),J_y(\br,t),J_z(\br,t))^T$ represents the source of
the electric field and $\mathcal H$ denotes the operator
\begin{equation}
{\mathcal H}  \equiv
\left( \begin{array}{cc} 0 &
  -\frac{1}{\sqrt{\mu}}\mathbf{\nabla}\times\frac{1}
  {\sqrt{\varepsilon}} \\
   \phantom{-}\frac{1}{\sqrt{\varepsilon}}\mathbf{\nabla}
   \times\frac{1}{\sqrt{\mu}} & 0
 \end{array}\right).
\label{Hconti}
\end{equation}
Writing $\bZ(t)=(\bX(t),\bY(t))^T$ it is easy to show that ${\mathcal H}$
is skew symmetric, i.e. ${\mathcal H}^T=-{\mathcal H}$, with respect to the
inner product $\langle\bZ(t)|\bZ^\prime(t)\rangle
\equiv\int_V\bZ^T(t)\cdot\bZ^\prime(t)\, d\br$, where $V$ denotes the system's
volume.
In addition to Eq.(\ref{TDMEJ}), the EM fields also satisfy
$\nabla\cdot(\sqrt{\mu}\bX(t)) = 0$ and
$\nabla\cdot(\sqrt{\ve}\bY(t)) = 0$~\cite{BornWolf}.
Throughout this paper we use dimensionless quantities: We measure distances in units of {$\lambda$}
and expresss time and frequency in units of {$\lambda/c$} and {$c/\lambda$}, respectively.

A numerical algorithm that solves the time-dependent Maxwell equations
necessarily involves some discretization procedure of the spatial derivatives
in Eq.~(\ref{TDMEJ}).
Ideally, this procedure should not change the basic symmetries of the Maxwell equations.
We will not discuss the (important) technicalities of the
spatial discretization (we refer the reader to Refs.~\cite{Taflove,Kunz})
as this is not essential to the discussion that follows.
On a spatial grid Maxwell's curl equations (\ref{TDMEJ})
can be written in the compact form~\cite{Kole01}
\begin{equation}
\dd{t}\bPsi(t)=H\bPsi(t)-\bPhi(t)\,.
\label{PsiJ}
\end{equation}
The vector $\bPsi(t)$ is a representation of $\bZ(t)$
on the grid. The matrix $H$ is the discrete analogue of the
operator (\ref{Hconti}), and the vector $\bPhi(t)$ contains all the
information on the current source $\bJ(t)$.
The formal solution of Eq.~(\ref{PsiJ}) is given by
\begin{equation}
\bPsi(t)=U(t)\bPsi(0)-\int_0^{t}U(t-u)\bPhi(u)du\,,
\label{FormalJ}
\end{equation}
where

\begin{equation}
U(t)=e^{tH},
\label{Ut}
\end{equation}
denotes the time-evolution matrix.
If the discretization procedure preserves the underlying symmetries of the time-dependent
Maxwell equations then the matrix $H$ is real and skew
symmetric~\cite{Kole01}, implying that $U(t)$ is orthogonal~\cite{WILKINSON}.
Physically, the orthogonality of $U(t)$ implies conservation of energy~\cite{Kole01}.

%-------------------------------------------------------------
%
\section{Time integration algorithms}\label{sec3}
%
%-------------------------------------------------------------
There are two, closely related, strategies to construct an algorithm for performing the time integration
of the time-dependent Maxwell equations defined on the grid~\cite{Smith85}.
The traditional approach is to discretize (with increasing level of sophistication)
the derivative with respect to time~\cite{Smith85}. The other is to approximate the formally exact solution, i.e.
the matrix exponential $U(t)=e^{tH}$ by some time evolution matrix $\widetilde U(t)$~\cite{Smith85,DeRaedt87}.
We adopt the latter approach in this paper as it facilitates the construction of algorithms with
specific features, such as unconditional stability~\cite{DeRaedt87,Kole01}.

If the approximation $\widetilde U(t)$ is itself an orthogonal transformation,
then $\Vert\widetilde U(t)\Vert=1$ where
$\Vert X\Vert$ denotes the 2-norm of a vector or matrix $X$~\cite{WILKINSON}.
In the absence of source terms (i.e. $\bPhi(t)=\bzero$) this implies that
$\Vert\widetilde\bPsi(t)\Vert=\Vert\widetilde{U}(t)\bPsi(0)\Vert = \Vert\bPsi(0)\Vert$,
for an arbitrary initial condition $\bPsi(0)$ and for all times $t$ and
hence the time integration algorithm defined by $\widetilde U(t)$
is unconditionally stable by construction~\cite{Smith85,DeRaedt87}.

In the presence of current sources, for general $\widetilde U(t)$,
it follows immediately from Eq.(\ref{FormalJ}) that

\begin{eqnarray}
\Vert \widetilde\bPsi(t)\Vert &\le&
\Vert\bPsi(t)\Vert + \widetilde\epsilon\left(\Vert\bPsi(0)\Vert+\int_0^t\Vert \bPhi(u)\Vert du\right),
\label{bound2}
\end{eqnarray}
where $\Vert\widetilde U(u)-U(u)\Vert\le\widetilde\epsilon$ for $0\le u\le t$
and $\widetilde\epsilon$ is a measure for
the accuracy of the approximation $\widetilde U(t)$.

From Eq.(\ref{FormalJ}) it follows that the EM fields $\bPsi(t)$ change according to

\begin{equation}
\bPsi(t+\tau)=e^{\tau H}\bPsi(t)-\int_t^{t+\tau}e^{(t+\tau-u)H}\bPhi(u)du.
\label{timestep0}
\end{equation}
In the time-stepping approach we approximate the source term in Eq.(\ref{timestep0}) by
the standard 3-point Gauss-Legendre quadrature formula~\cite{ABRAMOWITZ}

\begin{equation}
\bPsi(t+\tau)=e^{\tau H}\bPsi(t)+\frac{\tau}{2}
\sum_{i=0}^2 w_i e^{(1+x_i)\tau H/2}\bPhi(t+(1+x_i)\tau/2)+{\cal O}(\tau^7),
\label{timestep}
\end{equation}
where $x_0,x_1,x_2$ are the zeros of the Legendre polynomial $P_3(x)=x(5x^2-3)/2$
and $w_i=8/(1-x_i^2)(15x_i^2-3)^2$~\cite{ABRAMOWITZ}.
In practice we replace $e^{(1+x_i)\tau H/2}$ in Eq.(\ref{timestep}) by an
approximation $\widetilde U((1+x_i)\tau/2)$.

We now consider three options to construct the approximate time evolution matrix $\widetilde U(t)$.
We exclude from the discussion the exceptional cases for which
the matrix exponential $U(t)=e^{tH}$ can be calculated directly,
as these are usually of little relevance for realistic problems.
The first approach yields the conventional Yee algorithm, a higher-order generalization thereof, and
the unconditional schemes proposed in Ref.\cite{Kole01}.
The second option is to use rational approximations to the exponential, yielding the standard ADI methods.
Finally, the Chebyshev polynomial approximation to the matrix exponential is used to construct a one-step
algorithm.

\subsection{Suzuki product-formula approach}

A systematic approach to construct approximations to matrix
exponentials is to make use of the Lie-Trotter-Suzuki formula~\cite{Suzuki77,Trotter59}
\begin{equation}
e^{tH}=e^{t(H_1+\ldots+H_p)}=
\lim_{m\rightarrow\infty}
\left(\prod_{i=1}^p e^{t{H}_i/m}\right)^m,
\label{TROT}
\end{equation}
and generalizations thereof~\cite{Suzuki8591,DeRaedt83}.
Expression Eq.~(\ref{TROT}) suggests that
\begin{equation}
U_1(\tau)=e^{\tau{H}_1}\ldots e^{\tau{H}_p},
\label{tsapprox}
\end{equation}
might be a good approximation to $U(\tau)$ if $\tau$ is sufficiently small.
Applied to the case of interest here, if all the $H_i$ are real and skew-symmetric $U_1(\tau)$
is orthogonal by construction and a numerical scheme based on
Eq.~(\ref{tsapprox}) will be unconditionally stable.
For orthogonal matrices $U(\tau)$ and $U_1(\tau)$ it can be shown that~\cite{DeRaedt87}
\begin{equation}
\|U(\tau)-U_1(\tau)\|\leq\frac{\tau^2}{2}\sum_{i<j}\|[{H}_i,{H}_j]\|\,,
\label{tserror}
\end{equation}
where $[{H}_i,{H}_j]=H_i H_j - H_j H_i$ is, in general, non-zero.
Relaxing the condition that $U(\tau)$ and $U_1(\tau)$ are orthogonal matrices
changes the $\tau$
dependence in Eq.~(\ref{tserror}) but for small $\tau$ the error still vanishes
like $\tau^2$~\cite{Suzuki8591}.
From Eq.~(\ref{tserror}) it follows that, in general, the Taylor series of
$U(\tau)$ and $U_1(\tau)$ are identical up to first order in $\tau$.
We will call $U_1(\tau)$ a first-order approximation to $U(\tau)$.

The product-formula approach provides simple, systematic procedures to
improve the accuracy of the approximation to $U(\tau)$ without changing its
fundamental symmetries.
For example the matrix
\begin{equation}
U_2(\tau)={U_1(-\tau/2)}^TU_1(\tau/2)=
e^{\tau{H}_p/2}\ldots e^{\tau{H}_1/2}e^{\tau{H}_1/2}\ldots e^{\tau{H}_p/2},
\label{secordapp}
\end{equation}
is a second-order approximation to $U(\tau)$~\cite{Suzuki8591,DeRaedt83}.
If $U_1(\tau)$ is orthogonal, so is $U_2(\tau)$.
For orthogonal $U_2(\tau)$ we have

\begin{equation}
\|U(t)-[U_2(t/m)]^m\|\leq c_2 \tau^2t,
\label{tserror2}
\end{equation}
where $c_2$ is a positive constant~\cite{DeRaedt87}.

Suzuki's fractal decomposition approach~\cite{Suzuki8591} gives a general
method to construct higher-order approximations based on $U_2(\tau)$ (or $U_1(\tau)$).
A particularly useful fourth-order approximation is given by~\cite{Suzuki8591}
\begin{equation}
U_4(\tau)=U_2(a\tau)U_2(a\tau)U_2((1-4a)\tau)U_2(a\tau)U_2(a\tau),
\label{fouordapp}
\end{equation}
where $a=1/(4-4^{1/3})$.
The approximations Eqs.(\ref{tsapprox}) and (\ref{secordapp}), and
(\ref{fouordapp}) have proven to be very useful in many applications
~\cite{Suzuki77,Chorin78,DeRaedt83,DeRaedt87,Koboyashi94,DeRaedt94,Rouhi95,Shadwick97,Krech98,Tran98,Michielsen98,DeRaedt00}
and, as we show below, turn out to be equally useful for solving the time-dependent Maxwell equations.
As before, for orthogonal $U_4(\tau)$ we have~\cite{DeRaedt87}

\begin{equation}
\|U(t)-[U_4(t/m)]^m\|\leq c_4 \tau^4 t,
\label{tserror4}
\end{equation}
where $c_4$ is a positive constant.

As our numerical results (see below) show,
for sufficiently small $\tau$, the numerical error of a time integrator
vanishes with $\tau$ according to the $\tau$-dependence of the corresponding
rigorous bound, e.g. Eqs.(\ref{tserror}),(\ref{tserror2}), or (\ref{tserror4}).
Our experience shows that if this behavior is not observed, there is a fair chance that
the program contains one or more errors.

In practice an efficient implementation of the first-order scheme is all
that is needed to construct the higher-order algorithms
Eqs.(\ref{secordapp}) and (\ref{fouordapp}).
The crucial step of this approach is to choose the $H_i$'s such that the matrix exponentials
$\exp(\tau H_1)$, ..., $\exp(\tau H_p)$ can be calculated efficiently.
This will turn the formal expressions for $U_2(\tau)$ and $U_4(\tau)$ into efficient
algorithms to solve the time-dependent Maxwell equations.

\subsection{ADI algorithms}

Instead of hunting for a decomposition that leads to matrix exponentials
$\exp(\tau H_1)$, ..., $\exp(\tau H_p)$ that are easy to compute, one can opt for
an algorithm in which each of these exponentials is calculated approximately.
In principle this might be beneficial because there is more flexibility in decomposing $H$.
The standard strategy, preserving the symmetry of $H_1$, ..., $H_p$
is to use rational (Pad\'e) approximations to the exponential~\cite{Smith85}.
For instance, the approximation $e^x\approx(1+x/2)/(1-x/2)$
with some decompostion $H=H_1+H_2$ yields the second-order-accurate
ADI algorithm~\cite{Smith85,NumericalRecipes,Horvath}

\begin{equation}
U_2^{ADI}(\tau)=(I-\tau H_1/2)^{-1}(I+\tau H_2/2)(I-\tau H_2/2)^{-1}(I+\tau H_1/2),
\label{ADI}
\end{equation}
where $I$ is the identity matrix.
As the subscript indicates, the algorithm (\ref{ADI}) is second-order accurate in time.
For general skew-symmetrix $H_1$ and $H_2$, it is easy to show that
the algorithm $U_2^{ADI}(\tau)$ is unconditionally stable.
Following Ref.~\cite{Horvath} we rearrange factors and obtain

\begin{eqnarray}
\|\left[U_2^{ADI}(\tau)\right]^m\|&=&\|(I-\tau H_1/2)^{-1}X_2 X_1 X_2 \ldots X_1 X_2(I+\tau H_1/2)\|
\nonumber \\
&\le& \|(I-\tau H_1/2)^{-1}\|\|X_2 X_1 X_2 \ldots X_1 X_2\|\|(I+\tau H_1/2)\|
=\|(I-\tau H_1/2)^{-1}\|\|(I+\tau H_1/2)\|.
\label{ADIproof}
\end{eqnarray}
We used the fact that for skew-symmetric
$H_i$, $X_i\equiv (I-\tau H_i/2)^{-1}(I+\tau H_i/2)$ is orthogonal and that
$\|X_2 X_1 X_2 \ldots X_1 X_2\|^2=1$.
If $X$ is skew-symmetric, it's eigenvalues are pure imaginary and therefore $(I-X)^{-1}$ is non-singular.
Hence, for any number of time steps $m$,
$\|\left[U_2^{ADI}(\tau)\right]^m\|\le C$, where $C$ is some
finite positive constant, proving that the $U_2^{ADI}(\tau)$ is unconditionally stable in the Lax-Richtmyer
sense~\cite{Smith85}.

The matrix inversions appearing in Eq.(\ref{ADI}) suggest that for practical purposes
the implicit method $U_2^{ADI}(\tau)$ will not be very useful unless $I-\tau H_1/2$ and $I-\tau H_2/2$
take special forms that allow efficient matrix inversion~\cite{Smith85,NumericalRecipes}.

\begin{figure}[t]
\begin{center}
\includegraphics[width=8.cm]{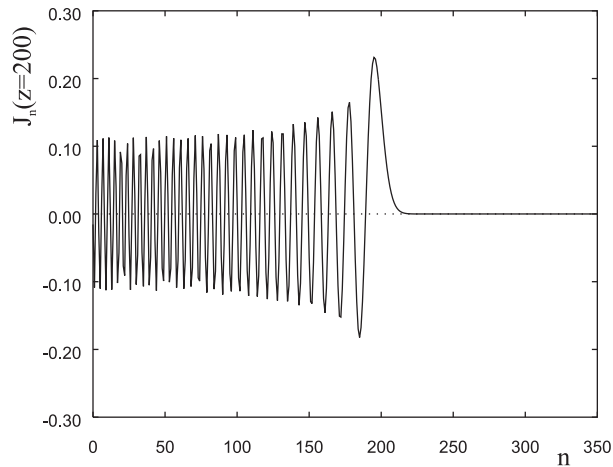}
\caption{Dependence of the Bessel function $J_n(z=200)$ on the order $n$.}
\label{fig:Jn}
\end{center}
\end{figure}

\subsection{One-step algorithm}

The basic idea of this approach is to make use of extremely accurate polynomial
approximations to the matrix exponential.
First we use the Chebyshev polynomial expansion
to approximate $U(t)$ and then show how to treat the source term
in Eq.~(\ref{FormalJ}).
We begin by ``normalizing'' the matrix $H$.
The eigenvalues of the skew-symmetric matrix $H$ are pure imaginary numbers.
In practice $H$ is sparse so it is easy to compute $\Vert H\Vert_1\equiv\max_j \sum_i |H_{i,j}|$.
Then, by construction, the eigenvalues of $B\equiv -iH/\Vert H\Vert_1$
all lie in the interval $[-1,1]$~\cite{WILKINSON}.
Expanding the initial value $\bPsi(0)$ in the (unknown) eigenvectors ${\bf b}_j$ of
$B$, we find from Eq.~(\ref{FormalJ}) with $\bPhi(t)\equiv 0$:
\begin{equation}
\bPsi(t)=
e^{izB}\bPsi(0) =\sum_j e^{izb_j} {\bf b}_j
\langle{\bf b}_j|\bPsi(0)\rangle,
\label{expz}
\end{equation}
where $z=t\Vert H\Vert_1$ and
the $b_j$ denote the (unknown) eigenvalues of $B$.
There is no need to know the eigenvalues and eigenvectors of $B$ explicitly.
We find the Chebyshev polynomial expansion of $U(t)$ by computing the
expansion coefficients of each of the functions $e^{izb_j}$ that appear
in Eq.~(\ref{expz}).
In particular, as $-1\le b_j \le 1$, we can use the expansion~\cite{ABRAMOWITZ}
$
e^{izb_j}=J_0(z) + 2\sum_{k=1}^{\infty} i^{k}J_{k}(z)T_{k}(b_j)\,,
$
where $J_k(z)$ is the Bessel function of integer order $k$
to write Eq.~(\ref{expz}) as
\begin{equation}
\bPsi(t)
=\left[J_0(z)I + 2\sum_{k=1}^{\infty} J_{k}(z)\hat{T}_{k}(B)\right] \bPsi(0)\,.
\label{SUM0}
\end{equation}
Here $\hat{T}_{k}(B)=i^k T_k(B)$ is a
matrix-valued modified Chebyshev polynomial that is defined by the recursion
relations
\begin{equation}
\hat{T}_{0}(B)\bPsi(0)=\bPsi(0)\label{CHEB1}\,,\quad
\hat{T}_{1}(B)\bPsi(0)=iB\bPsi(0)\,,
\label{CHEB44}
\end{equation}
and
\begin{equation}
\hat{T}_{k+1}(B)\bPsi(0)=
2iB\hat{T}_{k}(B)\bPsi(0)+\hat{T}_{k-1}(B)\bPsi(0)\,,
\label{CHEB4}
\end{equation}
for $k\ge1$.

As $\Vert \hat{T}_{k}(B) \Vert\le1$ by construction and
$|J_k(z)|\le |z|^k/2^k k!$ for $z$ real~\cite{ABRAMOWITZ},
the resulting error vanishes exponentially fast for sufficiently large $K$.
Thus, we can obtain an accurate approximation by summing
contributions in Eq.~(\ref{SUM0}) with $k\leq K$ only.
The number $K$ is fixed by requiring that $|J_{k}(z)|<\kappa$ for all $k>K$.
Here, $\kappa$ is a control parameter that determines the accuracy of the approximation.
For fixed $\kappa$, $K$ increases linearly with $z=t\Vert H\Vert_1$ (there is
no requirement on $t$ being small).
From numerical analysis it is known that for fixed $K$, the Chebyshev polynomial is very nearly
the same polynomial as the minimax polynomial \cite{NumericalRecipes}, i.e.
the polynomial of degree $K$ that has the smallest maximum deviation from the true function,
and is much more accurate than for instance a Taylor expansion of the same degree $K$.
In Fig.\ref{fig:Jn} we show a plot of $J_n(z=200)$ as a function of $n$
to illustrate these points.
From Fig.\ref{fig:Jn} it is clear that the Chebyshev polynomial expansion will only be useful if
$K$ lies to the right of the right-most extremum of $J_n(z=200)$, i.e. $K$ has to be larger than
200 in this example.

We now turn to the treatment of the current source $\bJ(t)$.
The contribution of the source term to the EM field at time $t$ is
given by the last term in Eq.~(\ref{FormalJ}).
For simplicity we only consider the case of a sinusoidal source
\begin{equation}
\bJ(\br,t)=\Theta(t)\Theta(T-t)\bj(\br)\sin(\Omega t),
\label{SOURCE}
\end{equation}
where $\bj(\br)$ specifies the spatial distribution and $\Omega\equiv 2\pi f_s$ the angular
frequency of the source. The step functions $\Theta(t)$ and $\Theta(T-t)$ indicate that
the source is turned on at $t=0$ and is switched off at $t=T$.
%Artifacts that result from the discontinuity at $t=T$ can be miminized by
%chosing $T$ such that $T\Omega/2\pi$ is an integer number.
Note that Eq.~(\ref{SOURCE}) may be used to compose sources with a more
complicated time dependence by a Fourier sine transformation.

The formal expression for the contribution of the sinusoidal source (\ref{SOURCE}) reads
\begin{eqnarray}
\int_0^t e^{(t-u){H}}\bPhi(u)\,du&=&
(\Omega^2+H^2)^{-1} e^{(t-T^\prime)H}
\times(\Omega e^{T^\prime H}-\Omega\cos\Omega T^\prime -H\sin\Omega T^\prime)\bXi \nonumber \\
&\equiv& f(H,t,T^\prime,\Omega)\bXi\,,
\label{FORMALC}
\end{eqnarray}
where $T^\prime=\min(t,T)$ and $\bPhi(u)\equiv\Theta(t)\Theta(T-t)\sin(\Omega t)\bXi$.
The vector $\bXi$ represents the spatial (time-independent) distribution $\bj(\br)$ and has
the same dimension as $\bPsi(0)$.
The coefficients of the Chebyshev polynomial expansion of
the formal solution (\ref{FORMALC}) are calculated as follows.
First we repeat the scaling procedure described above and substitute in
Eq.~(\ref{FORMALC}) $H=ix\Vert H\Vert_1$, $t=z/\Vert H\Vert_1$,
$T^\prime=Z^\prime/\Vert H\Vert_1$, and $\Omega=\omega\Vert H\Vert_1$.
Then, we compute the (Fast) Fourier Transform with respect to $x$ of the function
$f(x,z,Z^\prime,\omega)$ (which is non-singular on the interval $-1\le x \le 1$).
By construction, the Fourier coefficients $S_k(t{\Vert H\Vert_1})$ are
the coefficients of the Chebyshev polynomial expansion~\cite{ABRAMOWITZ}.

Taking into account all contributions of the source term with $k$ smaller than
$K^\prime$ (determined by a procedure similar to the one for $K$),
the one-step algorithm to compute the EM fields at time $t$ reads
\begin{eqnarray}
\hat{\bPsi}(t)&=&
\left[J_0(t{\Vert H\Vert_1})I +
2\sum_{k=1}^{K} J_{k}(t{\Vert H\Vert_1})\hat{T}_{k}(B)\right] \bPsi(0)
+%\nonumber\\&+&
\left[S_0(t{\Vert H\Vert_1})I + 2\sum_{k=1}^{K^\prime} S_{k}(t{\Vert H\Vert_1})
\hat{T}_{k}(B)\right] \bXi\,.
\label{APPROXJ}
\end{eqnarray}
Note that in this one-step approach the time dependence of
the source is taken into account exactly, without actually sampling it.

In a strict sense, the one-step method does not yield an orthogonal approximation.
However, for practical purposes it can be viewed as an extremely stable
time-integration algorithm because it yields an approximation to the
exact time evolution operator $U(t)=e^{tH}$ that is exact to nearly machine precision, i.e.
in practice the value of $\widetilde\epsilon$ in Eq.(\ref{bound2}) is very small.
This also implies that within the same precision
$\nabla\cdot(\mu\bH(t)) = \nabla\cdot(\mu\bH(t = 0))$ and
$\nabla\cdot(\ve\bE(t)) = \nabla\cdot(\ve\bE(t = 0))$ holds for all
times, implying that the numerical scheme will
not produce artificial charges during the time integration~\cite{Taflove,Kunz}.

%-------------------------------------------------------------
%
\section{Implementation}\label{sec4}
%
%-------------------------------------------------------------
The basic steps in the construction of the product-formula and one-step
algorithms are best illustrated by considering the simplest case, i.e. the Maxwell
equations of a 1D homogeneous problem. From a conceptual point of view nothing is lost by doing this:
the extension to 2D and 3D nonhomogeneous problems is straigthforward,
albeit technically non-trivial~\cite{Kole01,Kole02,hdrcheb0,hdrcheb1}.

We consider a system, infinitely large in the $y$ and $z$ direction, for which $\ve=1$ and $\mu=1$.
Under these conditions, the Maxwell equations reduce to two independent sets
of first-order differential equations~\cite{BornWolf}, the transverse electric (TE) mode
and the transverse magnetic (TM) mode~\cite{BornWolf}.
As the equations of the TE- and TM-mode differ by a sign we can restrict our considerations to the TM-mode only.
The magnetic field $H_y(x,t)$ and the electric field $E_z(x,t)$
of the TM-mode in the 1D cavity of length $L$ are solutions of
\begin{eqnarray}
\dd{t}H_y(x,t) & = & \dd{x}E_z(x,t),
\label{TMH}\\
\dd{t}E_z(x,t) & = & \dd{x}H_y(x,t) - J_z(x,t),
\label{TME}
\end{eqnarray}
subject to the boundary condition $E_z(0,t)=E_z(L,t)=0$~\cite{BornWolf}.
Note that the divergence of both fields are trivially zero.

\begin{figure}[t]
\begin{center}
\includegraphics[width=8.cm]{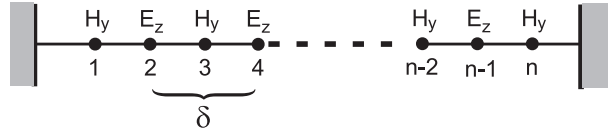}
\caption{Positions of the two TM-mode field components on the one-dimensional grid.
The distance between two next-nearest neighbors is denoted by $\delta$.}
\label{fig:fig1}
\end{center}
\end{figure}

Following Yee~\cite{Yee66}, to discretize Eqs.(\ref{TMH}) and (\ref{TME}),
it is convenient to assign $H_y$ to odd and $E_z$ to even numbered lattice sites,
as shown in Fig.~\ref{fig:fig1}.
Using the second-order central-difference approximation to the first
derivative with respect to $x$, we obtain

\begin{eqnarray}
\dd{t}H_y(2i+1,t) & = &\delta^{-1} (E_z(2i+2,t)-E_z(2i,t)),
\label{eqn:discrTMx} \\
\dd{t}E_z(2i,t) & = &\delta^{-1} (H_y(2i+1,t)-H_y(2i-1,t)) - J_z(2i,t),
\label{eqn:discrTMy}
\end{eqnarray}
where we have introduced the notation $A(i,t)=A(x=i\delta/2,t)$.
The integer $i$ labels the grid points and $\delta$ denotes the
distance between two next-nearest neighbors on the lattice (hence the
absence of a factor two in the nominator).
We define the $n$-dimensional vector $\bPsi(t)$ by
\begin{equation}
\bPsi(i,t) =
\left\{ \begin{array}{lll} H_y(i,t), & \mbox{$i$ odd}
\\ E_z(i,t), & \mbox{$i$ even}
\end{array} \right..
\label{eqn:eqnindx}
\end{equation}
The vector $\bPsi(t)$ contains both the magnetic and the electric field
on the lattice points $i=1,\ldots,n$.
The $i$-th element of $\bPsi(t)$ is given by the inner product
$\bPsi(i,t)=\be^T_i\cdot\bPsi(t)$ where $\be_i$ denotes the $i$-th unit vector
in the $n$-dimensional vector space.
Using this notation (which proves most useful for the case of 2D and
3D for which it is rather cumbersome to write down explicit matrix representations),
it is easy to show that Eqs.(\ref{eqn:discrTMx}) and
(\ref{eqn:discrTMy}) can be written in the form (\ref{PsiJ})
where the matrix $H$ is given by
\begin{equation}
H=
\delta^{-1}
\sum_{i=1}^{n-1}
\left(
\be^{\pht}_{i}\be^T_{i+1}-
\be^{\pht}_{i+1}\be^T_{i}\right)
=
\left( \begin{array}{ccccc}
0 & \delta^{-1} & & & \\
-\delta^{-1} & 0 & \delta^{-1} & & \\
& \ddots & \ddots & \ddots & \\
& & -\delta^{-1} & 0 & \delta^{-1} \\
& & & -\delta^{-1} & 0
\end{array} \right).
\label{H}
\label{matrixH}
\end{equation}
We immediately see that $H$ is sparse and skew-symmetric by construction.

\subsection{Yee-type algorithms}

First we demonstrate that the Yee algorithm fits into the product-formula approach.
For the 1D model (\ref{matrixH}) it is easy to see that one
time-step with the Yee algorithm corresponds
to the operation

\begin{eqnarray}
U_1^{Yee}(\tau)=(I+\tau A)(I-\tau A^T)=e^{\tau A} e^{-\tau A^T},
\label{U1Yee}
\end{eqnarray}
where

\begin{eqnarray}
A &=&
\delta^{-1}
\mathop{{\sum}'}_{i=2}^{n-1}
\left(
\be^{\pht}_{i}\be^T_{i-1}-
\be^{\pht}_{i}\be^T_{i+1}
\right)
=
\left( \begin{array}{cccccc}
0  &  \delta^{-1} & 0 &           0 & 0 &\cdots\\
0  &            0 & 0 &           0 & 0 &\cdots\\
0  & -\delta^{-1} & 0 & \delta^{-1} & 0 &\cdots\\
0  &            0 & 0 &           0 & 0 &\cdots\\
0  &            0 & 0 & -\delta^{-1}& 0 &\cdots \\
\vdots&    \vdots & \vdots & \vdots& \vdots & \ddots \\
\end{array} \right),
\label{A}
\end{eqnarray}
and we used the arrangements of $H$ and $E$ fields as defined by Eq.(\ref{eqn:eqnindx}).
We use the notation $\mathop{{\sum}'}$ to indicate that the stride of the summation index is two.
Note that since $A^2=0$ we have $e^{\tau A}=1+\tau A$ {\sl exactly}.
Therefore we recover the time-step operator of the Yee algorithm
using the first-order product formula approximation to $e^{\tau H}$
and decomposing $H=A-A^T$.
However, the Yee algorithm is second-order, not first order, accurate in time~\cite{Taflove,Kunz}.
This is due to the use of a staggered grid in time~\cite{Taflove,Kunz}.
To perform one time step with the Yee algorithm we need to know
the values of $E_z(t)$ and $H_y(t+\tau/2)$, not $H_y(t)$.
Another method has to supply the $H_y$-field at a time shifted by $\tau/2$.

Within the spirit of this approach, we can easily eliminate
the staggered-in-time grid at virtually no extra computational
cost or progamming effort (if a conventional Yee code is available)
by using the second-order product formula

\begin{eqnarray}
U_2^{Yee}(\tau)=e^{\tau A/2} e^{-\tau A^T} e^{\tau A/2} = (I+\tau A/2)(I-\tau A^T)(I+\tau A/2).
\label{Yee2}
\end{eqnarray}
The effect of the last factor is to propagate the $H_y$-field by $\tau/2$. The middle factor
propagates the $E_z$-field by $\tau$. The first factor again propagates the $H_y$ field by $\tau/2$.
In this scheme all EM fields are to be taken at the same time.
The algorithm defined by $U_2^{Yee}(\tau)$ is second-order accurate in time
by construction~\cite{DeRaedt87}.
Note that $e^{\tau A/2}$ is not orthogonal so nothing has been gained in terms of stability.
Since $\left(U_2^{Yee}(\tau)\right)^m=e^{-\tau A/2}\left(U_1^{Yee}(\tau)\right)^{m}e^{+\tau A/2}$,
we see that, compared to the original Yee algorithm, the extra computational work is proportional
to $(1+2/m)$, hence negligible if the number of time steps $m$ is large.

According to the general theory outlined in Sec.\ref{sec3}, the expression

\begin{eqnarray}
U_4^{Yee}(\tau)=U_2^{Yee}(a\tau)U_2^{Yee}(a\tau)U_2^{Yee}((1-4a)\tau)U_2^{Yee}(a\tau)U_2^{Yee}(a\tau),
\end{eqnarray}
defines a fourth-order accurate Yee-like scheme, the realization of which requires almost no effort
once $U_2^{Yee}$ has been implemented.
It is easy to see that the above construction of the Yee-like algorithms
holds for the much more complicated 2D, and 3D inhomogeneous case as well.
Also note that the fourth-order Yee algorithm $U_4^{Yee}$ does not require extra storage to
hold field values at intermediate times.

\subsection{Unconditionally stable algorithms}

Guided by previous work on Schr\"odinger and diffusion
problems~\cite{DeRaedt87,DeRaedt94,Michielsen98}, we split $H$ into two parts
\begin{eqnarray}
H_{1} &=&
\delta^{-1}
\mathop{{\sum}'}_{i=1}^{n-1}
\left(
\be^{\pht}_{i}\be^T_{i+1}-
\be^{\pht}_{i+1}\be^T_{i}
\right)
=
\left( \begin{array}{cccccc}
0  & \delta^{-1} & 0 & 0 & 0 &\cdots\\
-\delta^{-1} & 0 & 0 & 0 & 0 &\cdots\\
0  & 0 & 0 & \delta^{-1} & 0 &\cdots\\
0  & 0 & -\delta^{-1} & 0 & 0 & \cdots\\
0  & 0 & 0 & 0 & 0 &\cdots \\
0  & 0 & 0 & 0 & -\delta^{-1} &\cdots \\
\vdots&\vdots & \vdots & \vdots& \vdots & \ddots \\
\end{array} \right),
\label{h1}
\\
H_{2} &=&
\delta^{-1}
\mathop{{\sum}'}_{i=1}^{n-2}
\left(
\be^{\pht}_{i+1}\be^T_{i+2}-
\be^{\pht}_{i+2}\be^T_{i+1}
\right)
=
\left( \begin{array}{cccccc}
0  &  0 & 0 & 0 & 0 &\cdots\\
0  &  0 & \delta^{-1} & 0 & 0 &\cdots\\
0  & -\delta^{-1} & 0 & 0 & 0 &\cdots\\
0  &  0 & 0 & 0 & \delta^{-1} &\cdots\\
0  &  0 & 0 & -\delta^{-1} & 0 &\cdots \\
0  &  0 & 0 & 0 & 0 &\cdots \\
\vdots&\vdots &\vdots& \vdots & \vdots & \ddots \\
\end{array} \right).
\label{h2}
\end{eqnarray}
such that $H=H_{1}+H_{2}$. In other words
we divide the lattice into odd and even numbered cells.
Clearly both $H_{1}$ and $H_{2}$ are skew-symmetric block-diagonal matrices,
containing one $1\times1$ matrix and $(n-1)/2$ real, $2\times2$
skew-symmetric matrices.
According to the general theory given above, the first-order algorithm is given by
\begin{eqnarray}
\widetilde U_1(\tau)&=&e^{\tau H_{1}} e^{\tau H_{2}}
%\nonumber \\&=&
=
\left\{\mathop{{\prod}'}_{i=1}^{n-1}
\exp\left[\tau\delta^{-1}\left(
\be^{\pht}_{i}\be^T_{i+1}-
\be^{\pht}_{i+1}\be^T_{i}
\right)\right]\right\}
\left\{\mathop{{\prod}'}_{i=1}^{n-2}
\exp\left[\tau\delta^{-1}\left(
\be^{\pht}_{i+1}\be^T_{i+2}-
\be^{\pht}_{i+2}\be^T_{i+1}
\right)\right]\right\}.
\label{U1D}
\end{eqnarray}
To derive Eq.(\ref{U1D}) we used the block-diagonal structure of $H_1$ and $H_2$
(see Eqs.(\ref{h1}) and (\ref{h2}))
and obtained an exact expression for $\widetilde U_1(\tau)$ in terms of an ordered
product of matrix exponentials: the order of the matrix exponentials between each pair
of curly brackets is irrelevant as these matrices commute with each other.
Each of these matrix exponentials only operates on a pair of elements of $\bPsi(t)$
and leaves other elements intact.
The indices of each of these pairs are given by the subscripts of $\be$ and
$\be^T$.
From Eq.(\ref{U1D}) it is clear what a program should do:
Make loops over $i$ with stride 2.
For each $i$ pick a pair of elements from $\Psi(t)$ according to the
subscripts of $\be$ and $\be^T$, compute (or fetch from memory) the elements of the plane rotation
(see Eq.~(\ref{twobytwo})), perform the plane rotation, i.e. multiply the
$2\times2$ matrices and the vectors of length two, and overwrite the same
two elements.
As the matrix exponential of a block-diagonal matrix is equal to the
block-diagonal matrix of the matrix exponentials of the individual blocks,
the numerical calculation of $e^{\tau H_{1}}$ (or $e^{\tau H_{2}}$) reduces
to the calculation of $(n-1)/2$ matrix exponentials of $2\times2$ matrices.
The matrix exponential of a typical $2\times2$ matrix appearing in
$e^{\tau H_{1}}$ or $e^{\tau H_{2}}$ is given by
\begin{eqnarray}
% \exp\left[\alpha\left(
% \be^{\pht}_{i}\be^T_{j}-
% \be^{\pht}_{j}\be^T_{i}
% \right)\right]
% \left(\begin{array}{c} \Psi(i,t) \\ \Psi(j,t) \end{array} \right)
% &=&
\exp\left[\alpha
\left(\begin{array}{cc} \php0&1\\ -1&0\end{array}\right)\right]
\left(\begin{array}{c} \Psi(i,t) \\ \Psi(j,t) \end{array} \right)
\label{twobytwo} % \\
&=&\left(
\begin{array}{cc}
\php\cos \alpha & \php\sin\alpha
\\ -\sin \alpha & \php\cos\alpha
\end{array}\right)
\left(\begin{array}{c} \Psi(i,t) \\ \Psi(j,t) \end{array} \right).
\end{eqnarray}
Using the algorithm to compute Eq.(\ref{U1D}), it is easy to construct the unconditionally stable,
higher-order algorithms
$\widetilde U_2(\tau)$ and $\widetilde U_4(\tau)$, see Eq.(\ref{secordapp}) and Eq.(\ref{fouordapp}).

Obviously, the decomposition into $H_1$ Eq.(\ref{h1}) and $H_2$ Eq.(\ref{h2}) yields the most
simple real-space algorithm. It is not difficult to imagine that a better but slightly more complicated
algorithm can be constructed by using blocks of $3\times3$ instead of $2\times2$ matrices.
Thus we are lead to consider the decomposition

\begin{eqnarray}
H_{3} &=&
\delta^{-1}
\mathop{{\sum}''}_{i=1}^{n-2}
\left(
\be^{\pht}_{i}\be^T_{i+1}
+\be^{\pht}_{i+1}\be^T_{i+2}
-\be^{\pht}_{i+1}\be^T_{i}
-\be^{\pht}_{i+2}\be^T_{i+1}
\right)
=
\left( \begin{array}{cccccc}
0  & \phantom{-}\delta^{-1} & 0 & 0 & 0 &\cdots\\
-\delta^{-1} & 0 & \phantom{-}\delta^{-1}  & 0 & 0 &\cdots\\
0  & -\delta^{-1}& 0  & 0 & 0 &\cdots\\
0  & 0 & 0 & 0 & 0 & \cdots\\
0  & 0 & 0 & 0 & 0 &\cdots \\
0  & 0 & 0 & 0 & -\delta^{-1}  &\cdots \\
\vdots&\vdots & \vdots & \vdots& \vdots & \ddots \\
\end{array} \right),
\label{h3}
\\
H_{4} &=&
\delta^{-1}
\mathop{{\sum}''}_{i=1}^{n-4}
\left(
\be^{\pht}_{i+2}\be^T_{i+3}
+\be^{\pht}_{i+3}\be^T_{i+4}
-\be^{\pht}_{i+3}\be^T_{i+2}
-\be^{\pht}_{i+4}\be^T_{i+3}
\right)
=
\left( \begin{array}{cccccc}
0  &  0 & 0 & 0 & 0 &\cdots\\
0  &  0 & 0 & 0 & 0 &\cdots\\
0  &  0 & 0 & \phantom{-}\delta^{-1} & 0 &\cdots\\
0  &  0 & -\delta^{-1}& 0 & \phantom{-}\delta^{-1} &\cdots\\
0  &  0 & 0 & -\delta^{-1} & 0 &\cdots \\
0  &  0 & 0 & 0 & 0  &\cdots \\
\vdots&\vdots &\vdots& \vdots & \vdots & \ddots \\
\end{array} \right).
\label{h4}
\end{eqnarray}
where the double prime indicated that the stride of the index $i$ is three.
Obviously both $H_{3}$ and $H_{4}$ are skew-symmetric block-diagonal matrices,
build from the $3\times3$ skew-symmetric matrix

\begin{eqnarray}
B &=&
\left( \begin{array}{ccc}
0  & \phantom{-}\delta^{-1} & 0 \\
-\delta^{-1} & 0 & \phantom{-}\delta^{-1} \\
0  & -\delta^{-1}& 0  \\
\end{array} \right).
\label{B}
\end{eqnarray}
As $B^3=-2B$ we have

\begin{eqnarray}
e^{\tau B}&=&1+sB+cB^2 % B^2 (1-\cos(\sqrt{2}\tau))/2+B\sin(\sqrt{2}\tau)
=
\left( \begin{array}{ccc}
1-c  & s & c \\
-s & 1-2c & s \\
-c  & -s& 1-c  \\
\end{array} \right),
\label{B2}
\end{eqnarray}
where $s=\sin(\sqrt{2}\tau)$ and $c=\sin^2(\tau/\sqrt{2})$.
In practice, using the $3\times3$ instead of the $2\times2$ decomposition is marginally more difficult.
We will denote the corresponding
second and fourth-order algorithm by $U_2^{3\times3}$ and $U_4^{3\times3}$ respectively.

\subsection{ADI algorithm}

For the tri-diagonal matrix (\ref{H}), the ADI algorithm reduces to
the Cranck-Nicholson method~\cite{NumericalRecipes}.
The tri-diagonal structure of the matrix $H$ permits the
calculation of $(I-\tau H/2)^{-1}\bPsi$ in ${\cal O}(n)$ operations by
standard linear algebra methods~\cite{NumericalRecipes}.

\subsection{One-step algorithm}

The one-step algorithm is based on the recursion (see Eqs.(\ref{CHEB44}) and (\ref{CHEB4}))

\begin{equation}
\bPsi_{k+1}=\frac{2H}{\Vert H\Vert_1}\bPsi_k+\bPsi_{k-1}.
\label{RECURSION}
\end{equation}
Thus, the explicit form Eq.(\ref{matrixH}) is all we need to implement the
matrix-vector operation (i.e. $\bPsi^\prime\leftarrow H\bPsi$) that enters Eq.(\ref{RECURSION}).

The coefficients $J_{k}(z)$ and $S_k(t)$ (see Eq.(\ref{APPROXJ}))
should be calculated to high precision.
Using the recursion relation of the Bessel functions, all $K$ coefficients
can be obtained with ${\cal O}(K)$ arithmetic operations \cite{NumericalRecipes}.
The numbers $S_k(t)$ can be calculated in ${\cal O}(K\log K)$ by standard Fast Fourier
transformation techniques.
Clearly both computations are a neglible fraction of the total computational cost
for solving the Maxwell equations.

Performing one time step amounts to repeatedly using recursion (\ref{CHEB4}) to
obtain $\hat{T}_{k}(B)\bPsi(0)$ for $k=2,\ldots,K$, multiply the elements
of this vector by $J_{k}(z)$ (or $S_k(z)$) and add all contributions.
This procedure requires storage for two vectors of the same length as $\bPsi(0)$
and some code to multiply such a vector by the sparse matrix $H$.
The result of performing one time step yields the solution at time $t$, hence
the name one-step algorithm.
In contrast to what Eqs.~(\ref{CHEB44}) and ~(\ref{CHEB4}) might suggest,
the algorithm does not require the use of complex arithmetic.

In the sequel, the caret $\hat{\phantom{\bPsi}} $ on top of a symbol indicates
that the results have been obtained by means of the one-step algorithm.
%-------------------------------------------------------------
%
\section{Numerical experiments}\label{sec5}
%
%-------------------------------------------------------------
\begin{figure}[t]
\begin{center}
\includegraphics[width=8.cm]{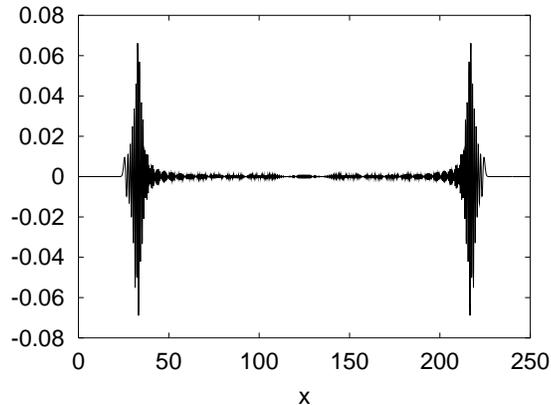}
\caption{The field $H_y(x,t=100)$ generated by a current source at $x=125$
that oscillates at frequency $f_s=1$ during the interval $0\le t\le 6$,
as obtained by the one-step algorithm %(\ref{APPROXJ})
with $K^\prime=2103$ ($K=0$ in this case).
}
\label{J1D1}
\end{center}
\end{figure}
\newcommand{\EXPON}[1]{E#1}
\begin{table}[t]
\begin{center}
\caption{%
The error $\Vert\tilde\bPsi(t) - \hat\bPsi(t) \Vert/\Vert\hat\bPsi(t) \Vert$
at time $t=100$ as a function of the time step $\tau$ for eight different FDTD algorithms.
The current source is positioned at the center of the system (see Fig.\ref{J1D1}),
and oscillates at frequency $f_s=1$ during the interval $0\le t\le 6$,
$\hat\bPsi(t)$ is the vector obtained by the one-step algorithm with $\kappa=10^{-9}$,
using $K^\prime=2103$ matrix-vector operations $\bPsi^\prime\leftarrow M\bPsi$.
Yee: $\tilde\bPsi(t)$ obtained by the Yee algorithm~\cite{Yee66,Taflove,Kunz};
Other columns: $\tilde\bPsi(t)$ obtained by the algorithms indicated.
}
\label{tab1}
\begin{ruledtabular}\begin{tabular}{ccccccccc}
$\tau$ & Yee & $U_2^{Yee}$ & $U_2^{ADI}$ & $\widetilde U_2$ & $\widetilde U_2^{3\times3}$ & $U_4^{Yee}$ & $\widetilde U_4$ & $\widetilde U_4^{3\times3}$  \\
\hline
\noalign{\vskip 4pt}
%  0.10000E+00  0.15907E+01  0.14872E+01  0.14646E+01  0.14666E+01  0.13738E+01  0.15097E+00  0.36772E+00  0.27065E+00
%  0.50000E-01  0.18452E+01  0.17902E+01  0.13056E+01  0.15669E+01  0.12785E+01  0.35581E-01  0.32709E-01  0.22797E-01
%  0.25000E-01  0.68358E+00  0.64476E+00  0.11720E+01  0.71810E+00  0.11575E+01  0.24878E-02  0.21889E-02  0.15129E-02
%  0.12500E-01  0.20752E+00  0.18909E+00  0.34591E+00  0.12950E+01  0.31123E+00  0.15754E-03  0.13911E-03  0.95954E-04
%  0.62500E-02  0.63547E-01  0.54493E-01  0.10494E+00  0.35134E+00  0.78284E-01  0.98770E-05  0.87304E-05  0.60201E-05
%  0.31250E-02  0.18677E-01  0.13802E-01  0.27532E-01  0.88261E-01  0.19586E-01  0.61777E-06  0.54622E-06  0.37665E-06
%  0.15625E-02  0.61429E-02  0.34534E-02  0.69055E-02  0.22071E-01  0.48970E-02  0.38618E-07  0.34148E-07  0.23549E-07
%  0.78125E-03  0.23400E-02  0.86339E-03  0.17267E-02  0.55178E-02  0.12243E-02  0.24213E-08  0.21431E-08  0.14846E-08
%  0.39063E-03  0.10069E-02  0.21585E-03  0.43169E-03  0.13795E-02  0.30607E-03  0.24440E-09  0.23841E-09  0.21608E-09
% & & & & & \\
  $0.10000\EXPON{+0}$& $0.16\EXPON{+1}$& $0.15\EXPON{+1}$& $0.15\EXPON{+1}$&  $0.15\EXPON{+1}$&  $0.14\EXPON{+1}$&  $0.15\EXPON{+0}$& $0.37\EXPON{+0}$&   $0.27\EXPON{+0}$\\
  $0.50000\EXPON{-1}$& $0.18\EXPON{+1}$& $0.18\EXPON{+1}$& $0.13\EXPON{+1}$&  $0.16\EXPON{+1}$&  $0.13\EXPON{+1}$&  $0.36\EXPON{-1}$& $0.33\EXPON{-1}$&   $0.23\EXPON{-1}$\\
  $0.25000\EXPON{-1}$& $0.68\EXPON{+0}$& $0.65\EXPON{+0}$& $0.12\EXPON{+1}$&  $0.72\EXPON{+0}$&  $0.12\EXPON{+1}$&  $0.25\EXPON{-2}$& $0.22\EXPON{-2}$&   $0.15\EXPON{-2}$\\
  $0.12500\EXPON{-1}$& $0.21\EXPON{+0}$& $0.19\EXPON{+0}$& $0.35\EXPON{+0}$&  $0.13\EXPON{+1}$&  $0.31\EXPON{+0}$&  $0.16\EXPON{-3}$& $0.14\EXPON{-3}$&   $0.96\EXPON{-4}$\\
  $0.62500\EXPON{-2}$& $0.63\EXPON{-1}$& $0.55\EXPON{-1}$& $0.10\EXPON{+0}$&  $0.35\EXPON{+0}$&  $0.78\EXPON{-1}$&  $0.99\EXPON{-5}$& $0.87\EXPON{-5}$&   $0.60\EXPON{-5}$\\
  $0.31250\EXPON{-2}$& $0.19\EXPON{-1}$& $0.14\EXPON{-1}$& $0.28\EXPON{-1}$&  $0.88\EXPON{-1}$&  $0.20\EXPON{-1}$&  $0.62\EXPON{-6}$& $0.55\EXPON{-6}$&   $0.38\EXPON{-6}$\\
  $0.15625\EXPON{-2}$& $0.61\EXPON{-2}$& $0.35\EXPON{-2}$& $0.69\EXPON{-1}$&  $0.22\EXPON{-1}$&  $0.49\EXPON{-2}$&  $0.39\EXPON{-7}$& $0.34\EXPON{-7}$&   $0.24\EXPON{-7}$\\
  $0.78125\EXPON{-3}$& $0.23\EXPON{-2}$& $0.86\EXPON{-2}$& $0.17\EXPON{-2}$&  $0.55\EXPON{-2}$&  $0.12\EXPON{-2}$&  $0.24\EXPON{-8}$& $0.21\EXPON{-8}$&   $0.15\EXPON{-8}$\\
  $0.39063\EXPON{-3}$& $0.10\EXPON{-2}$& $0.22\EXPON{-3}$& $0.43\EXPON{-3}$&  $0.14\EXPON{-2}$&  $0.31\EXPON{-3}$&  $0.24\EXPON{-9}$& $0.24\EXPON{-9}$&   $0.22\EXPON{-9}$
\end{tabular}
\end{ruledtabular}
\end{center}
\end{table}

\begin{figure}[t]
\includegraphics[width=12cm]{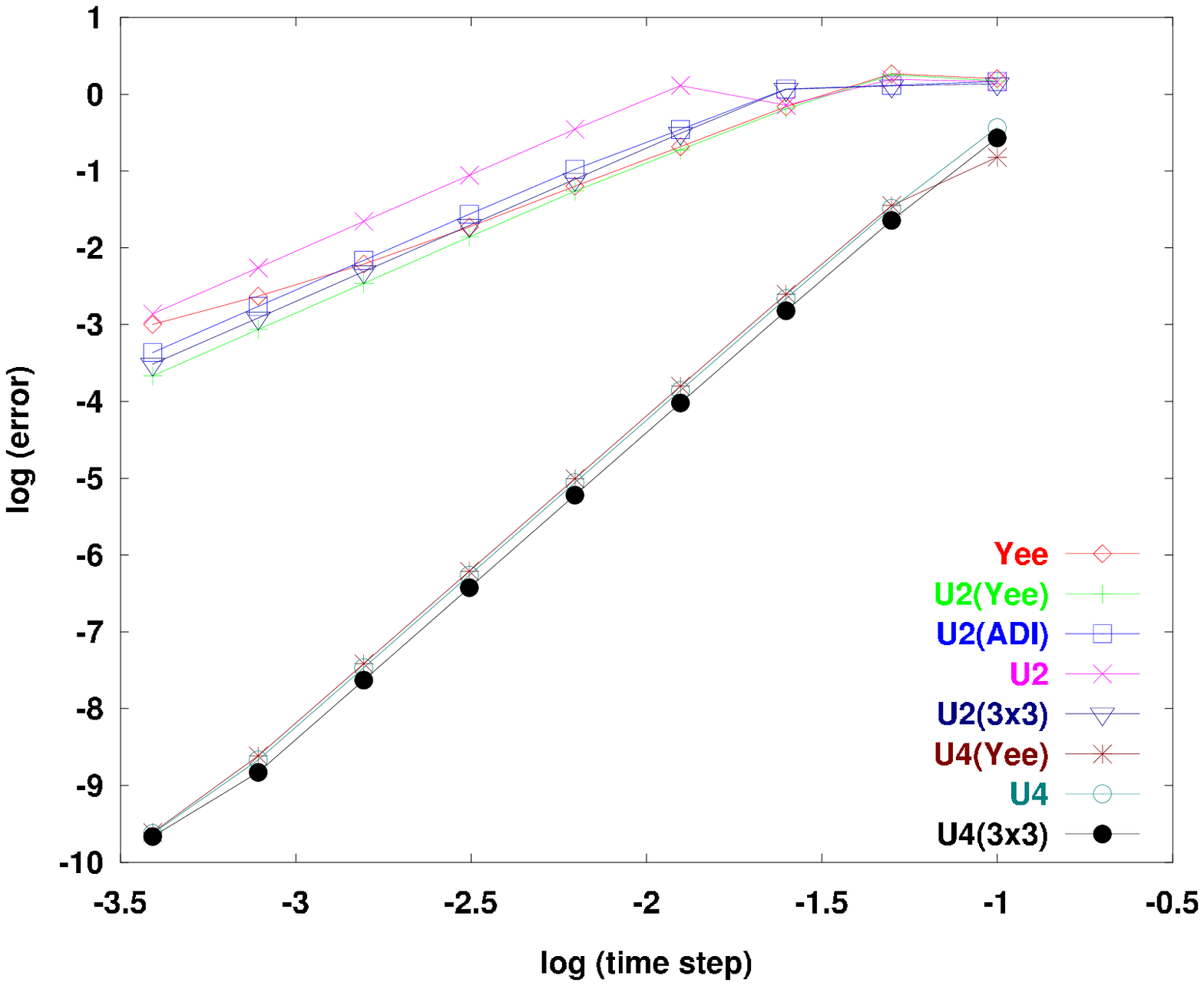}
\caption{%
The data presented in Table \ref{tab1} plotted on a double logarithmic scale.
Lines are guide to the eye.
}
\label{figtable1}
\end{figure}

\begin{table}[t]
\begin{center}
\caption{%
The error $\Vert\tilde\bPsi(t) - \hat\bPsi(t) \Vert/\Vert\hat\bPsi(t) \Vert$
at time $t=100$ as a function of the time step $\tau$ for eight different FDTD algorithms.
The system is the same as in Fig.\ref{J1D1} and Table \ref{tab1}.
The initial values of the EM fields are random, distributed uniformly
over the interval [-1,1].
$\hat\bPsi(t)$ is the vector obtained by the one-step algorithm $\kappa=10^{-9}$,
using $K=2080$ matrix-vector operations $\bPsi^\prime\leftarrow M\bPsi$.
Yee: $\tilde\bPsi(t)$ obtained by the Yee algorithm~\cite{Yee66,Taflove,Kunz};
Other columns: $\tilde\bPsi(t)$ obtained by the algorithms indicated.
}
\label{tab2}
\begin{ruledtabular}\begin{tabular}{ccccccccc}
$\tau$ & Yee & $U_2^{Yee}$ & $U_2^{ADI}$ & $\widetilde U_2$ & $\widetilde U_2^{3\times3}$ & $U_4^{Yee}$ & $\widetilde U_4$ & $\widetilde U_4^{3\times3}$  \\
\hline
%   0.10000E+00  0.98529E+01  0.11138E+02  0.13907E+01  0.14452E+01  0.13644E+01  0.11253E+01  0.12975E+01  0.56491E+00
%   0.50000E-01  0.12602E+01  0.12967E+01  0.13280E+01  0.13113E+01  0.14274E+01  0.78072E+00  0.16428E+00  0.44406E-01
%   0.25000E-01  0.13089E+01  0.13209E+01  0.13177E+01  0.12643E+01  0.12359E+01  0.57039E-01  0.11163E-01  0.28876E-02
%   0.12500E-01  0.12413E+01  0.12442E+01  0.13652E+01  0.11518E+01  0.62714E+00  0.36205E-02  0.71136E-03  0.18222E-03
%   0.62500E-02  0.69628E+00  0.69676E+00  0.11873E+01  0.32379E+00  0.16406E+00  0.22710E-03  0.44674E-04  0.11416E-04
%   0.31250E-02  0.18264E+00  0.18266E+00  0.36148E+00  0.81512E-01  0.41126E-01  0.14207E-04  0.27955E-05  0.71394E-06
%   0.15625E-02  0.45790E-01  0.45789E-01  0.91501E-01  0.20386E-01  0.10283E-01  0.88811E-06  0.17477E-06  0.44628E-07
%   0.78125E-03  0.11449E-01  0.11449E-01  0.22895E-01  0.50964E-02  0.25707E-02  0.55510E-07  0.10924E-07  0.27895E-08
%   0.39063E-03  0.28623E-02  0.28621E-02  0.57241E-02  0.12741E-02  0.64267E-03  0.34694E-08  0.68446E-09  0.17808E-09
\noalign{\vskip 4pt}
  $0.10000\EXPON{+0}$& $0.99\EXPON{+1}$& $0.11\EXPON{+2}$&  $0.14\EXPON{+1}$&  $0.15\EXPON{+1}$&  $0.17\EXPON{+1}$& $0.11\EXPON{+1}$&  $0.13\EXPON{+1}$&  $0.13\EXPON{+1}$\\
  $0.50000\EXPON{-1}$& $0.13\EXPON{+1}$& $0.13\EXPON{+1}$&  $0.13\EXPON{+1}$&  $0.13\EXPON{+1}$&  $0.14\EXPON{+1}$& $0.78\EXPON{+0}$&  $0.16\EXPON{+0}$&  $0.16\EXPON{+0}$\\
  $0.25000\EXPON{-1}$& $0.13\EXPON{+1}$& $0.13\EXPON{+1}$&  $0.13\EXPON{+1}$&  $0.13\EXPON{+1}$&  $0.12\EXPON{+1}$& $0.57\EXPON{-2}$&  $0.11\EXPON{-1}$&  $0.11\EXPON{-1}$\\
  $0.12500\EXPON{-1}$& $0.12\EXPON{+1}$& $0.12\EXPON{+1}$&  $0.14\EXPON{+1}$&  $0.12\EXPON{+1}$&  $0.63\EXPON{+0}$& $0.36\EXPON{-2}$&  $0.71\EXPON{-3}$&  $0.71\EXPON{-3}$\\
  $0.62500\EXPON{-2}$& $0.70\EXPON{+0}$& $0.70\EXPON{+0}$&  $0.12\EXPON{+1}$&  $0.32\EXPON{+0}$&  $0.16\EXPON{+0}$& $0.22\EXPON{-3}$&  $0.45\EXPON{-4}$&  $0.45\EXPON{-4}$\\
  $0.31250\EXPON{-2}$& $0.18\EXPON{+0}$& $0.18\EXPON{+0}$&  $0.36\EXPON{+0}$&  $0.82\EXPON{-1}$&  $0.41\EXPON{-1}$& $0.14\EXPON{-4}$&  $0.28\EXPON{-5}$&  $0.28\EXPON{-5}$\\
  $0.15625\EXPON{-2}$& $0.46\EXPON{-1}$& $0.46\EXPON{-1}$&  $0.92\EXPON{-1}$&  $0.20\EXPON{-1}$&  $0.10\EXPON{-1}$& $0.89\EXPON{-6}$&  $0.17\EXPON{-6}$&  $0.17\EXPON{-6}$\\
  $0.78125\EXPON{-3}$& $0.11\EXPON{-1}$& $0.11\EXPON{-1}$&  $0.23\EXPON{-1}$&  $0.51\EXPON{-2}$&  $0.26\EXPON{-2}$& $0.56\EXPON{-7}$&  $0.11\EXPON{-7}$&  $0.28\EXPON{-8}$\\
  $0.39063\EXPON{-3}$& $0.29\EXPON{-2}$& $0.29\EXPON{-2}$&  $0.57\EXPON{-2}$&  $0.13\EXPON{-2}$&  $0.64\EXPON{-3}$& $0.35\EXPON{-8}$&  $0.68\EXPON{-9}$&  $0.18\EXPON{-9}$\\
\end{tabular}
\end{ruledtabular}
\end{center}
\end{table}

\begin{figure}[t]
\includegraphics[width=12cm]{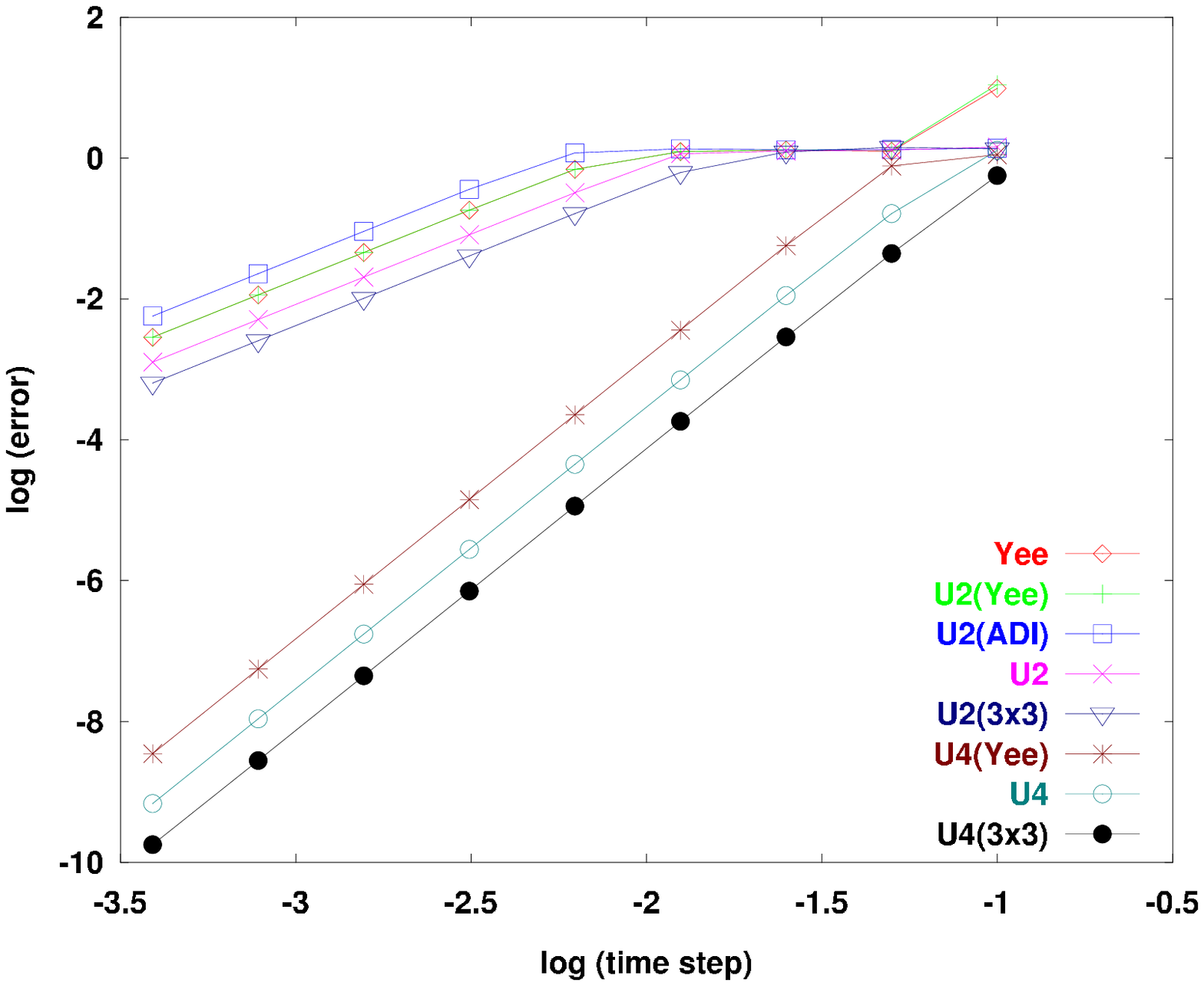}
\caption{%
The data presented in Table \ref{tab2} plotted on a double logarithmic scale.
Lines are guide to the eye.
}
\label{figtable2}
\end{figure}

\begin{table}[t]
\begin{center}
\caption{%
The error $\Vert\tilde\bPsi(t) - \hat\bPsi(t) \Vert/\Vert\hat\bPsi(t) \Vert$
at time $t=100$ as a function of the time step $\tau$ for eight different FDTD algorithms
The system is the same as in Fig.\ref{J1D1} and Table \ref{tab1}.
The initial state of the EM fields is a Gaussian wave packet
($E_z(t)=\exp(-(x-x_0-t)^2/\sigma^2$)
with a width $\sigma=4$, and its center $x_0=125$ positioned at the middle of the system (see Fig.\ref{J1D1}).
$\hat\bPsi(t)$ is the vector obtained by the one-step algorithm with $\kappa=10^{-9}$,
using $K=2080$ matrix-vector operations $\bPsi^\prime\leftarrow M\bPsi$.
Yee: $\tilde\bPsi(t)$ obtained by the Yee algorithm~\cite{Yee66,Taflove,Kunz};
Other columns: $\tilde\bPsi(t)$ obtained by the algorithms indicated.
}
\label{tab3}
\begin{ruledtabular}\begin{tabular}{ccccccccc}
$\tau$ & Yee & $U_2^{Yee}$ & $U_2^{ADI}$ & $\widetilde U_2$ & $\widetilde U_2^{3\times3}$ & $U_4^{Yee}$ & $\widetilde U_4$ & $\widetilde U_4^{3\times3}$  \\
\hline
%   0.10000E+00  0.25266E-02  0.25231E-02  0.50366E-02  0.14139E+01  0.79222E+00  0.27872E-06  0.15120E-01  0.17134E-01
%   0.50000E-01  0.63195E-03  0.63055E-03  0.12598E-02  0.90630E+00  0.25064E+00  0.17428E-07  0.95408E-03  0.14546E-02
%   0.25000E-01  0.15809E-03  0.15762E-03  0.31500E-03  0.25730E+00  0.64912E-01  0.10893E-08  0.59771E-04  0.96709E-04
%   0.12500E-01  0.39565E-04  0.39405E-04  0.78752E-04  0.64984E-01  0.16341E-01  0.68996E-10  0.37379E-05  0.61342E-05
%   0.62500E-02  0.99114E-05  0.98512E-05  0.19688E-04  0.16263E-01  0.40919E-02  0.12350E-10  0.23365E-06  0.38480E-06
%   0.31250E-02  0.24878E-05  0.24628E-05  0.49221E-05  0.40666E-02  0.10234E-02  0.11622E-10  0.14597E-07  0.24077E-07
%   0.15625E-02  0.62693E-06  0.61570E-06  0.12305E-05  0.10167E-02  0.25587E-03  0.11622E-10  0.90685E-09  0.15111E-08
%   0.78125E-03  0.15925E-06  0.15393E-06  0.30763E-06  0.25417E-03  0.63970E-04  0.11623E-10  0.55482E-10  0.10360E-09
%   0.39063E-03  0.41096E-07  0.38483E-07  0.76905E-07  0.63544E-04  0.15992E-04  0.11620E-10  0.43185E-10  0.46041E-10
\noalign{\vskip 4pt}
% & & & & & \\
  $0.10000\EXPON{+0}$& $0.25\EXPON{-2}$& $0.25\EXPON{-2}$&  $0.50\EXPON{-2}$&  $0.14\EXPON{+1}$&  $0.79\EXPON{+0}$& $0.28\EXPON{-6}$ &  $0.15\EXPON{-1}$ &  $0.17\EXPON{-1}$ \\
  $0.50000\EXPON{-1}$& $0.63\EXPON{-3}$& $0.63\EXPON{-3}$&  $0.13\EXPON{-2}$&  $0.90\EXPON{+0}$&  $0.25\EXPON{+0}$& $0.17\EXPON{-7}$ &  $0.95\EXPON{-3}$ &  $0.15\EXPON{-3}$ \\
  $0.25000\EXPON{-1}$& $0.16\EXPON{-3}$& $0.16\EXPON{-3}$&  $0.32\EXPON{-3}$&  $0.26\EXPON{+0}$&  $0.65\EXPON{-1}$& $0.11\EXPON{-8}$ &  $0.60\EXPON{-4}$ &  $0.97\EXPON{-4}$ \\
  $0.12500\EXPON{-1}$& $0.40\EXPON{-4}$& $0.39\EXPON{-4}$&  $0.79\EXPON{-4}$&  $0.65\EXPON{-1}$&  $0.16\EXPON{-1}$& $0.69\EXPON{-10}$&  $0.37\EXPON{-5}$ &  $0.61\EXPON{-5}$ \\
  $0.62500\EXPON{-2}$& $0.99\EXPON{-5}$& $0.98\EXPON{-5}$&  $0.20\EXPON{-4}$&  $0.16\EXPON{-1}$&  $0.41\EXPON{-2}$& $0.12\EXPON{-10}$&  $0.23\EXPON{-6}$ &  $0.38\EXPON{-6}$ \\
  $0.31250\EXPON{-2}$& $0.25\EXPON{-5}$& $0.25\EXPON{-5}$&  $0.49\EXPON{-5}$&  $0.41\EXPON{-2}$&  $0.10\EXPON{-2}$& $0.12\EXPON{-10}$&  $0.15\EXPON{-7}$ &  $0.24\EXPON{-7}$ \\
  $0.15625\EXPON{-2}$& $0.63\EXPON{-6}$& $0.61\EXPON{-6}$&  $0.12\EXPON{-5}$&  $0.10\EXPON{-2}$&  $0.26\EXPON{-3}$& $0.12\EXPON{-10}$&  $0.91\EXPON{-9}$ &  $0.15\EXPON{-8}$ \\
  $0.78125\EXPON{-3}$& $0.16\EXPON{-6}$& $0.15\EXPON{-6}$&  $0.31\EXPON{-6}$&  $0.25\EXPON{-3}$&  $0.64\EXPON{-4}$& $0.12\EXPON{-10}$&  $0.55\EXPON{-10}$&  $0.10\EXPON{-9}$\\
  $0.39063\EXPON{-3}$& $0.41\EXPON{-7}$& $0.38\EXPON{-7}$&  $0.77\EXPON{-7}$&  $0.64\EXPON{-4}$&  $0.16\EXPON{-4}$& $0.12\EXPON{-10}$&  $0.43\EXPON{-10}$&  $0.46\EXPON{-10}$\\
\end{tabular}
\end{ruledtabular}
\end{center}
\end{table}

\begin{figure}[t]
\includegraphics[width=12cm]{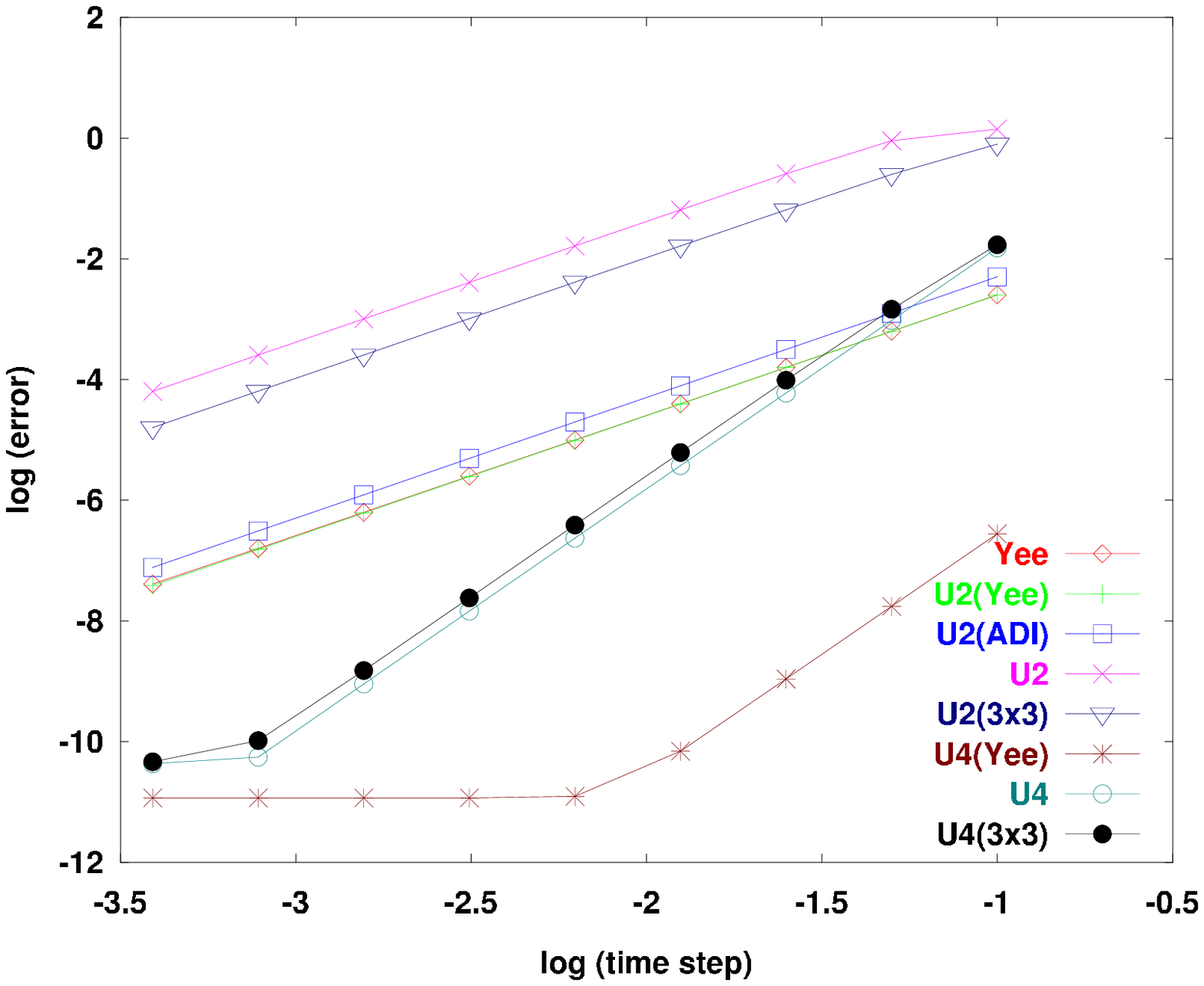}
\caption{%
The data presented in Table \ref{tab3} plotted on a double logarithmic scale.
Lines are guide to the eye.
}
\label{figtable3}
\end{figure}

Except for the conventional Yee algorithm, all algorithms discussed in this paper
operate on the vector of fields defined at the same time $t$.
We use the one-step algorithm (with a time step $\tau/2$) to compute
$E_z(\tau/2)$ and $H_y(\tau/2)$. Then we use $E_z(0)$ and $H_y(\tau/2)$ as
the initial values for the Yee algorithm.
In the presence of a current source, there are some ambiguities with this procedure
as it is not obvious how to treat the source term in Eq.(\ref{timestep}).
In order to permit a comparison of the final result of the conventional Yee algorithm
with those of the other methods, we use the one-step algorithm once more to shift the time
of the $H_y$ field by $-\tau/2$.
This procedure to prepare the initial and to analyse the final state of the Yee algorithm
does in fact make the results of the Yee algorithm look a little more accurate than
they would be if the exact data of the $\tau/2$-shifted fields are
not available. Thus, the results on the errors of the conventional Yee algorithm presented
in this paper give a too optimistic view on the accuracy of this algorithm
but we nevertheless adopt the above procedure
to make a quantitative comparison between the various algorithms.

We define the error of the solution $\tilde\bPsi(t)$ for the wave form by
$\Vert\tilde\bPsi(t) - \hat\bPsi(t) \Vert/\Vert\hat\bPsi(t) \Vert$
where $\hat\bPsi(t)$ is the vector of EM fields obtained by the one-step algorithm.
Thereby we have already assumed that the one-step algorithm yields the
exact (within numerical precision) results but this has to be demonstrated of course.
A comparison of the results of an unconditionally stable algorithm, e.g. $\widetilde U_4$
with those of the one-step algorithm is sufficient to show that
within rounding errors the latter yields the exact answer.
Using the triangle inequality

\begin{equation}
\Vert\bPsi(t) - \hat{\bPsi}(t) \Vert
\le
\Vert\bPsi(t) - \tilde\bPsi(t) \Vert +
\Vert\tilde\bPsi(t) - \hat{\bPsi}(t) \Vert,
\label{triangle}
\end{equation}
and the rigorous bound

\begin{equation}
\Vert\bPsi(t) - \tilde\bPsi(t) \Vert\le c_4\tau^4 t
\left(\Vert\bPsi(0)\Vert+\int_0^t\Vert \bJ(u)\Vert du\right),
\label{U4bound}
\end{equation}
we can be confident that the one-step algorithm yields the numerically exact answer if i)
Eq.(\ref{U4bound}) is not violated and ii) if
$\Vert\tilde\bPsi(t) - \hat{\bPsi}(t) \Vert$ vanishes like $\tau^4$.

In Fig.\ref{J1D1} we show a typical result of a one-step calculation
on a grid of $n=5001$ sites with $\delta=0.1$
(corresponding to a physical length of $250.05$),
and a current source placed at $i=2500$
to eliminate possible artifacts of the boundaries.
The frequency of the source is set to one ($f_s=1$) and the number of periods the source
radiates is set to six (i.e. $T=6$).
In Table \ref{tab1} (Fig.\ref{figtable1}) we present results for the errors, as
obtained by repeating the simulation shown in Fig.\ref{J1D1}
using eight different FDTD methods.
In Tables \ref{tab2} and \ref{tab3} (Figs.\ref{figtable2} and \ref{figtable3} respectively)
we shown similar results but instead
of using a current source, a random wave form (Table \ref{tab2})
and Gaussian wave packet (Table \ref{tab3}) was taken as the initial condition.

From the data in Tables \ref{tab1}, \ref{tab2} and \ref{tab3} we conclude that
the error of algorithm $\widetilde U_4$ vanishes like $\tau^4$, demonstrating that the one-step algorithm
yields the numerically exact result (see Eqs.\ref{triangle} and \ref{U4bound}).
The results presented in Tables \ref{tab2} and \ref{tab3} have been obtained
by using a vector of initial values that is normalized to one, i.e. $\|\bPsi(0)\|=1$.
As $\|\hat\bPsi(t)\|=1$ for $0\le t \le 100$ to at least 10 digits,
$\Vert\tilde\bPsi(t) - \hat\bPsi(t) \Vert/\Vert\hat\bPsi(t) \Vert=\Vert\tilde\bPsi(t) - \hat\bPsi(t) \Vert$
for all entries in Tables \ref{tab2} and \ref{tab3}.
The high precision of the one-step algorithm also allows us to use it for genuine time stepping
with arbitrarily large time steps, this in spite of the fact that strictly speaking,
the one-step algorithm is not unconditionally stable.

\begin{figure}[t]
\begin{center}
\includegraphics[width=14cm]{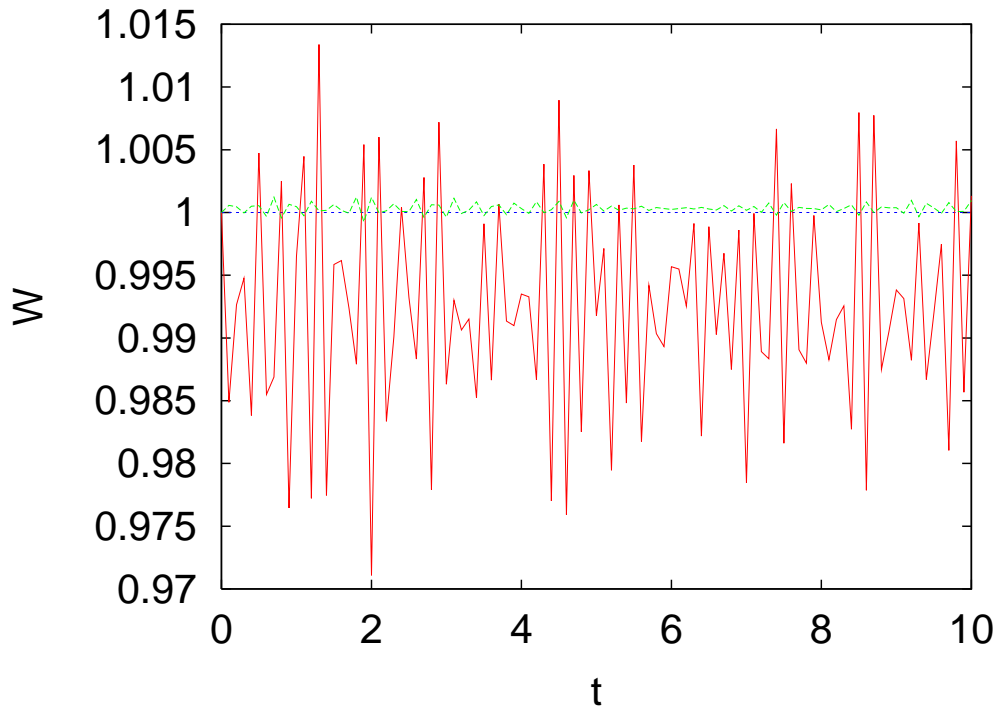}
\caption{
The energy of the EM fields $W=\bPsi^T(t)\cdot\bPsi(t)$ as the function of time
as obtained by the Yee (solid line), $U_2^{Yee}$ (dashed line), and $\widetilde U_4$ (dotted line)
algorithm for a 1D cavity
of size 48.05 ($n=97$ mesh points), a mesh size $\delta=0.1$ and a time step $\tau=0.01$.}
\label{energy}
\end{center}
\end{figure}

The data in Tables \ref{tab1}, \ref{tab2} and \ref{tab3}
%(and Figs.\ref{figtable1},\ref{figtable2}, and \ref{figtable3} respectively)
suggests that there does not seem to be a significant difference between the conventional
Yee algorithm and its variant $U_2^{Yee}$ but in fact there is.
The time evolution matrix corresponding to the Yee and the $U_2^{Yee}$ algorithm is not orthogonal.
Therefore the energy of the electromagnetic field ($\Vert\bPsi(t)\Vert^2$) is not conserved.
Furthermore, in the conventional Yee algorithm,
the $E$ and $H$-fields are time-shifted by $\tau/2$ with respect to each other.
These artifacts of are less prominent if we use $U_2^{Yee}$ instead of the Yee algorithm.
In Fig.\ref{energy} we show results of the time evolution of the total energy of the
EM field, for a system of n=97 sites, a mesh size $\delta=0.1$ and a time step of $\tau=0.01$.
A normalized ($\Vert\bPsi(0)\Vert^2=1$) random initial condition was used.
Furthermore, for this type of application, it makes no sense to invoke the
procedure described at the beginning of this section to time-shift one of the EM-fields by $\tau/2$:
as this operation has to be performed at each time step and is computationally expensive (because
it is numerically exact), we could as well use the same numerically exact procedure for time stepping
itself.

\begin{figure}[t]
\begin{center}
\includegraphics[width=14cm]{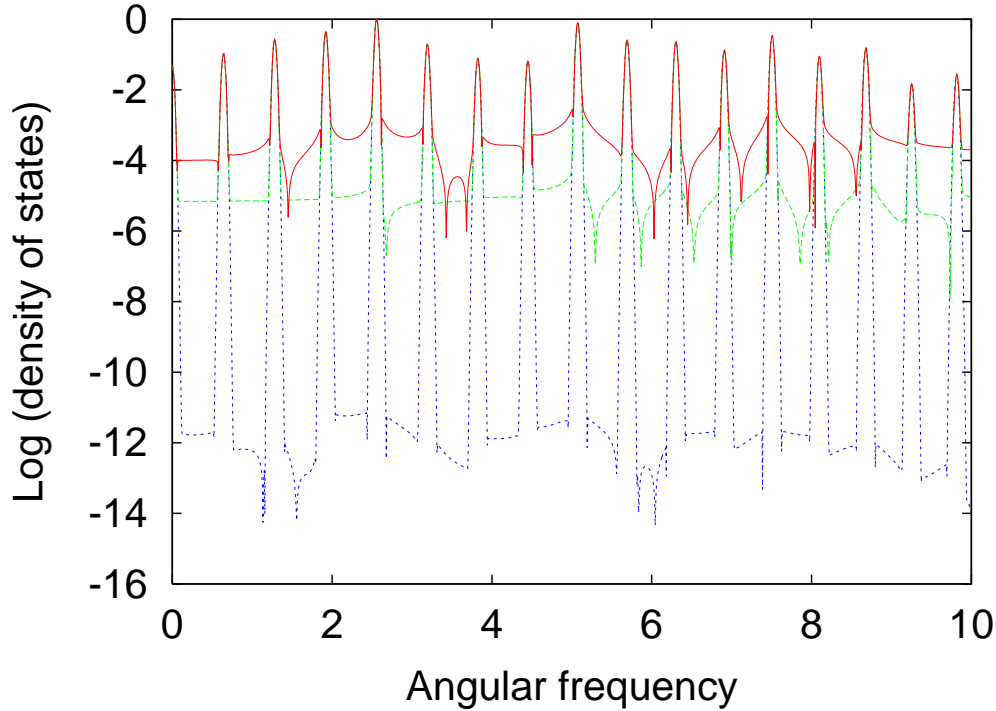}
\caption{The eigenvalues distribution of the matrix $H$, as
obtained by Fourier transformation of $\bPsi^T(t)\cdot\bPsi(0)$, for the
same system as in Fig.\ref{energy}.
The function $\bPsi^T(t)\cdot\bPsi(0)$ is sampled at time intervals of 0.1,
the total number of samples being 4096.
Solid line: Yee algorithm;
dashed line: $U_2^{Yee}$ algorithm;
dotted line: energy conserving algorithm $\widetilde U_4$.}
\label{spectrum}
\end{center}
\end{figure}

For the Yee algorithm, the fluctuations of the energy are a factor of ten larger than
in the case of the $U_2^{Yee}$ algorithm.
As expected on theoretical grounds, the $\widetilde U_4$ algorithm (dotted, horizontal line)
exactly conserves the energy.
The fact that $U_2^{Yee}$ conserves EM-field energy much better than the Yee algorithm also
has a considerable impact on the quality of the eigenmode distribution.
The latter is obtained by Fourier transformation of $\bPsi^T(t)\cdot\bPsi(0)$
(see ref.\cite{Kole01} for more details).
In Fig.\ref{spectrum} we show the low-frequency part of the eigenmode distribution
of the same system as the one of Fig.\ref{energy}.
It is obvious that there is a significant improvement in the quality of the
spectrum if we use $U_2^{Yee}$ instead of the Yee algorithm but for this
application $\widetilde U_4$ performs much better than the Yee-type algorithms.

%

% (disregarding the fact that the latter operates on the EM fields taken at the same time $t$).
Table \ref{tab1} suggests that \Two is the least efficient
of the five FDTD methods: It uses more arithmetic operations than the Yee algorithm
and yields errors that are larger than those of the Yee algorithm.
However, this conclusion is biased by the choice of the model problem and does not generalize.
If the initial EM field distribution is random then,
for sufficiently small $\tau$, algorithm \Two is more accurate
than the two second-order accurate Yee algorithms,
as is clear from the data in Table \ref{tab2}~\cite{explains}.
Also in this case, for the largest $\tau$ in Table \ref{tab2}, the Yee and $U_2^{Yee}$ algorithm
are operating at the point of instability, signaled by the fact that the norm of $\bPsi(t)$ grows
rapidly, resulting in errors that are very large.
From Tables \ref{tab1} and \ref{tab2} one might conclude that
the decomposition that generates Yee-type algorithms yields the least accurate
approximations to the time evolution operator, although the difference is not really significant,
but, as Table \ref{tab3} shows, this conclusion would be wrong.
If the initial state is a Gaussian wave packet that is fairly broad, the Yee-type algorithms
are much more accurate than the unconditionally stable algorithms employed in this paper.
From the data in Tables \ref{tab1}, \ref{tab2} and \ref{tab3} we conclude that there
is no good reason to use the ADI algorithm (even disregarding the fact that it is slower than
the other second-order methods).
In general the $\widetilde U_2^{3\times3}$ ($\widetilde U_4^{3\times3}$)
algorithm performs a little better than $\widetilde U_2$ ($\widetilde U_4$)
but the gain is marginal.
In contrast to the numerical data presented in Ref.\cite{Horvath}, for all algorithms
the data of Tables \ref{tab1},\ref{tab2} and \ref{tab3} clearly agree with
the theoretically expected behavior of the error as a function of $\tau$
if $\tau$ is small enough~\cite{exception}.

Usually if a current source is present we have $\bPsi(0)=\bzero$.
Then the one-step algorithm requires $K^\prime$ (sparse)
matrix-vector operations ($\bPsi^\prime\leftarrow M\bPsi$) to compute $\bPsi(t)$.
For a 1D system the standard Yee, $U_2^{Yee}$ and $U_4^{Yee}$ ,\Two , and \Tfour  algorithms
perform (in worst case, without additional optimization), respectively, 1, 3/2, 6, 3/2, and 6 $M\bPsi$-operations per time step.
The one-step algorithm carries out $K^\prime=2103$ matrix-vector operations $\bPsi^\prime\leftarrow M\bPsi$
to complete this simulation.
This implies that for all $\tau<t/K^\prime$, the FDTD algorithms will perform
more $\bPsi^\prime\leftarrow M\bPsi$ operations than the one-step algorithm.
For the data presented in this paper, this is the case if $\tau<0.05$ for the Yee algorithm and is always the case
for \Tfour  because the latter uses a factor of 6 more $\bPsi^\prime\leftarrow M\bPsi$ operations
than the Yee algorithm.

%-------------------------------------------------------------
%
\section{Conclusion}\label{sec6}
%
%-------------------------------------------------------------

The answer to the question which of the algorithms is the most efficient crucially one
depends on the accuracy that one finds acceptable.
Taking the data of Table \ref{tab1} as an example we see
that if one is satisfied with an error of more than 2\%, one could use the Yee algorithm.
With $\tau=0.05$ it needs 2000 time steps to find the solution at $t=100$, close to the $K^\prime=2103$.
Nevertheless we recommend to use the one-step algorithm because then the time-integration error is neglegible.
The Yee algorithm is no competition for \Tfour  if one requires an error of less than 1\%
but then \Tfour is not nearly as efficient (by a factor of about 6) as the one-step algorithm.
Increasing the dimensionality of the problem favors the one-step algorithm~\cite{hdrcheb0,hdrcheb1}.
These conclusions seem to be quite general
and are in concert with numerical experiments on 1D, 2D and 3D systems~\cite{hdrcheb1}.
A simple theoretical analysis of the $\tau$ dependence of the error shows that
the one-step algorithm is more efficient than any other FDTD method
if we are interested in the EM fields at a particular (large) time only~\cite{hdrcheb0,hdrcheb1}.
This may open possibilities to solve problems in computational electrodynamics that are currently intractable.
The Yee-like algorithms do not conserve the energy of the EM fields and therefore they are less suited
for the calculation of the eigenvalue distributions (density of states), a problem for which
the $\widetilde U_4$ algorithm may be the most efficient of all the algorithms covered in the paper.

The main limitation of the one-step algorithm lies in its mathematical justification.
The Chebyshev approach requires that $H$ is diagonalizable and that its eigenvalues
are real or pure imaginary.
The effect of relaxing these conditions on the applicability of the Chebyshev approach is left
for future research.

In this paper we have focused entirely on the accuracy of the time integration algorithms,
using the most simple discretization of the spatial derivatives.
In practice it is straightforward, though technically non-trivial, to treat
more sophisticated discretization schemes~\cite{Taflove,Kole02}
by the methodology reviewed is this paper.

\begin{acknowledgments}
H.D.R. and K.M. are grateful to T. Iitaka for drawing our attention to
the potential of the Chebyshev method and for illuminating discussions.
This work is partially supported by the Dutch `Stichting Nationale Computer Faciliteiten'
(NCF), and the EC IST project CANVAD.
\end{acknowledgments}


\begin{thebibliography}{99}
\bibitem{COMMENT1}
This is due to the specific split-up adopted in \cite{Harsh00}
and not an intrinsic property of the spectral-domain approach.
\bibitem{exception}
The data for small $\tau$, obtained from the Yee algorithm for the case of a current source is
an exception. This is due to the ambiguities mentioned in Sec.\ref{sec5}.
\end{thebibliography}
\end{document}